\documentclass[]{spie}  

\usepackage[]{graphicx}

\title{Realization and application of a 111 million pixel 
       backside-illuminated detector and camera} 

\author{Norbert Zacharias\supit{a}, Bryan Dorland\supit{a},
        Richard Bredthauer\supit{b}, Kasey Boggs\supit{b},
        Greg Bredthauer\supit{b}, and Mike Lesser\supit{c}
\skiplinehalf
\supit{a}United States Naval Observatory, 3450 Massachusetts Avenue NW, 
           Washington DC, 20392;\\
\supit{b}Semiconductor Technology Associates, 
           San Juan Capistrano, CA 92675;\\
\supit{c}Steward Observatory, University of Arizona, 
           Tucson, AZ 85721
}

\authorinfo{Further author information: (Send correspondence to N.Z.)\\
   E-mail: nz@usno.navy.mil, Telephone: 1 202 762 1423}

\begin{document} 
  \maketitle 


\begin{abstract}
  A full-wafer, 10,580 $\times$ 10,560 pixel (95 $\times$ 95 mm) 
  CCD was designed and tested
  at Semiconductor Technology Associates (STA) with 9 $\mu$m square 
  pixels and 16 outputs. The chip was successfully fabricated in 
  2006 at DALSA and some performance results are presented here.
  This program was funded by the Office of Naval Research through a
  Small Business Innovation in Research (SBIR) program requested by 
  the U.S.~Naval Observatory for its next generation
  astrometric sky survey programs.  
  Using Leach electronics, low read-noise output of the 111 million 
  pixels requires 16 seconds at 0.9 MHz.
  Alternative electronics developed at STA allow readout at 20 MHz.
  Some modifications of the design to include anti-blooming features, 
  a larger number of outputs, and use of p-channel material for space 
  applications are discussed.
\end{abstract}




\section{INTRODUCTION}

This paper descibes the motivations and requirements which
led to the development of the world's largest-format CCD detector.
The history and realization of the entire camera around this
device for the U.S.~Naval Observatory (USNO) is presented.
Some performance results obtained from a thinned, backside-illuminated 
detector of this kind are given.
The paper concludes with plans for the future and explains
applications in ground-based and space-based programs.

\section{REQUIREMENTS}

In recent years the requirements for ever larger focal plane
assemblies at astronomical telescopes have mostly been satisfied
by assembling numerous smaller devices into large focal plane
mosaics.  The advantage of this mosaic approach is a larger
yield in producing high quality detectors, which results in cost
savings for the instrument development.  Although larger,
monolithic detectors are desirable, there has not been a
real driver requiring the bold step to go beyond the current
typical 2k $\times$ 4k scale devices.

\subsection{Astrometry}

In planning projects beyond the successful USNO CCD Astrograph Catalog 
(UCAC) program\cite{ucac2} it was realized that a large-format detector
is needed.  Using the technique of photographic astrometry
the positions of stars are determined with respect to several
reference stars with known positions by direct imaging
of the sky using dedicated telescopes.   The large-format
photographic plates (up to about 17 inches on a side) traditionally 
used were later replaced by CCD detectors, providing higher quantum
efficiency and more accurate centroiding results as compared to
the photographic process.  Unfortunately, CCD
detectors are very small compared to photographic plates.
For high accuracy astrometric measurements many reference stars
need to be on the same detector or at least many well exposed
anonymous stars are required to tile together overlapping
fields with as few mapping parameters as possible.
We needed to advance beyond the existing CCD formats in order to 
make significant progress\cite{potsdam} in this area and to fully
utilize the existing large focal plane of our astrograph as well as
those of future dedicated astrometric telescopes.

At that same meeting\cite{potsdam} where the need for larger-format 
CCD detectors for astrometric mapping was presented, we learned
that designers of a Large Binocular Telescope spectroscopic 
instrument were also looking at similar types of CCD detectors 
to improve calibrations and lower systematic errors.
Following discussions, both groups agreed to share risks in 
development of a large-format detector which could be used for 
both projects.
A pixel size of 9 ${\mu}m$ was agreed upon, but independent
funding avenues had to be pursued.

Of particular importance for astrometric applications 
are a high charge transfer efficiency and relatively fast 
readout, with a goal of about 10 sec for the full frame. 
This required the use of a large number of parallel 
outputs. Standard high-quality materials and designs 
used for science grade CCD detectors were sufficient to
satisfy all other requirements, providing the yield issue
could be addressed successfully.

\subsection{USNO Telescopes}

Figure 1 shows the USNO Twin Astrograph, which was used for the
UCAC program (1997 to 2004) and before that for astrometry using
photographic plates (24 cm square and 8 $\times$ 10 inch).
The original ``blue" lens was replaced by a 5-lens ``red lens"
objective\cite{redlens} of extremely high astrometric performance, 
which has been in operation since 1990.
The new 10k camera dewar is attached to the red lens now,
while the second telescope tube features a visual bandpass
corrected lens which is used for guiding.

   \begin{figure}
   \begin{center}
   \begin{tabular}{c}
   \includegraphics[height=12cm]{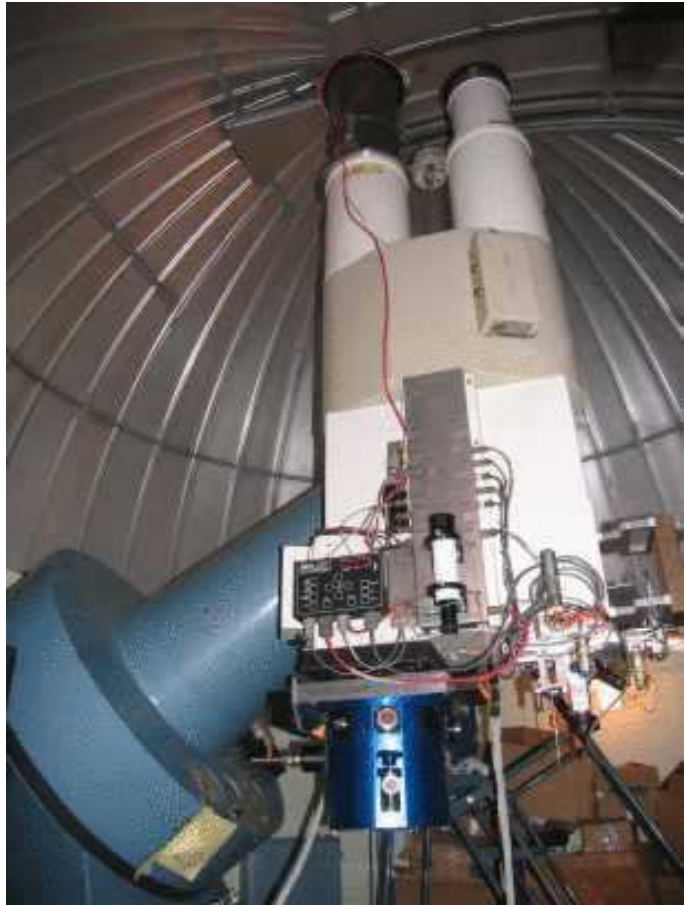}
   \end{tabular}
   \end{center}
   \caption[example] 
   { \label{fig:astrogr} 
     The U.S.~Naval Observatory Twin Astrograph, currently at the 
     Flagstaff Station, with 10k camera dewar attached to the 
    ``red lens."
     This setup will be used for astrometric test observations
     on the sky in preparation for the URAT program.}
   \end{figure} 

The following table presents the main characteristics of the
existing USNO astrograph and the planned USNO Robotic Astrometric 
Telescope (URAT)\cite{uratd,uratp}.
Both feature an available focal plane area of about 30 cm in diameter.
Design work on the URAT began in 2000 and concluded in 2005.
A contract was signed with EOST to produce the primary mirror
which will be delivered by the end of 2007.
Funding of the URAT telescope is uncertain beyond that time.
However, the focal plane development is progressing well
with major purchases anticipated in 2008.

\begin{table}[h]
\caption{Comparison of the existing USNO Twin astrograph (red lens)
   telescope with the planned USNO Robotic Astrometric Telescope (URAT).}
\label{tab:fonts}
\begin{center}       
\begin{tabular}{|l|c|c|} 
\hline
\rule[-1ex]{0pt}{3.5ex}  property      & astrograph &  URAT  \\
\hline
\rule[-1ex]{0pt}{3.5ex}  aperture [meter]         &  0.20   & 0.85  \\ 
\hline
\rule[-1ex]{0pt}{3.5ex}  focal length [meter]     &  2.00   & 3.60  \\
\hline
\rule[-1ex]{0pt}{3.5ex}  scale [arcsec/pixel]     &  0.90   & 0.50  \\
\hline
\rule[-1ex]{0pt}{3.5ex}  diameter field of view [degree] &  9.00 & 4.50  \\
\hline
\rule[-1ex]{0pt}{3.5ex}  diameter focal plane [mm]       &  320  & 283   \\
\hline
\rule[-1ex]{0pt}{3.5ex}  bandpass [nm]  & 550$-$750 & 600$-$800 \\
\hline
\end{tabular}
\end{center}
\end{table} 

The astrograph will serve as a testbed for any future URAT
focal plane assemblies, and might even be used for a new
all-sky survey (see below).
An upgrade of its control interface with integration into
the new camera system is in progress.

\section{REALIZATION}

\subsection{Research Program}

A sponsor for the development of a general, large-format, monolithic 
detector was found at the Office of Naval Research.
A Small Business Innovation in Research (SBIR) topic proposed by
the USNO Astometry Department was accepted for a phase I study
in 2004.
Originally an 8-inch full-wafer CMOS or CMOS/hybrid device was 
considered, but this was quickly rejected as unrealistic at the time.
Instead, 2 companies were funded in phase I to develop a 6-inch wafer 
full-frame CCD detector.
In 2005 the main research phase II funding was awarded to one of
the phase I participants,
Semiconductor Technology Associates (STA), of San Juan Capistrano, CA.

\subsection{Chip Design}

This 10,580 $\times$ 10,560 pixel array was designed and
developed at STA. 
A combination of new features went into the CCD design
to increase yield and lower risk and costs.

To achieve high frame rates it is necessary to clock both the parallel and
serial clocks at high rates.  Existing bussing for the serial registers is
satisfactory for transfers at rates in excess of 40 MHz.  A parallel clock
gate is normally bussed from the left and right sides with aluminum straps.
Resistance across the gate is usually only several thousand ohms, 
with a distributed capacitance of several picofarads.  It is normally
possible to clock these structures at several hundred kilohertz.   
However, the very
large size of the STA1600 yields parallel gates that are over 95 mm long,
with a resistance close to 790K ohms.  This permits only very slow parallel
clocking.  To reduce the effective gate resistance a metal grid is placed
over the image area.   Three micron metal lines run vertically and
horizontally over the polysilicon gates.  Periodic contacts connect to the
polysilicon, lowering its resistance.   The small metal lines require
contacts smaller than is possible with the 1x masks used for the CCD
manufacture.  To obtain the required 1 micron contacts we used a second
contact mask solely for this strapping.  These contacts are printed with a
photostepper and a 5x mask plate.   This mixed mask approach is simpler and
more cost effective for creating such a large device.  Using only a stepper
alone for manufacturing a CCD of this size would require many additional
masks and be much more costly.

\subsection{CCD Fabrication}

The STA1600 full-frame detector was successfully manufactured
by DALSA in June 2006 (see press release).
The yield was sufficient to produce several engineering and
science grade chips.
Initial characterization by STA of the full-wafer device
confirms acceptable parameters\cite{111mpx}.
Figure 2 shows the packaged device with 4 connectors for 
4 outputs each.
 
   \begin{figure}
   \begin{center}
   \begin{tabular}{c}
   \includegraphics[height=10cm]{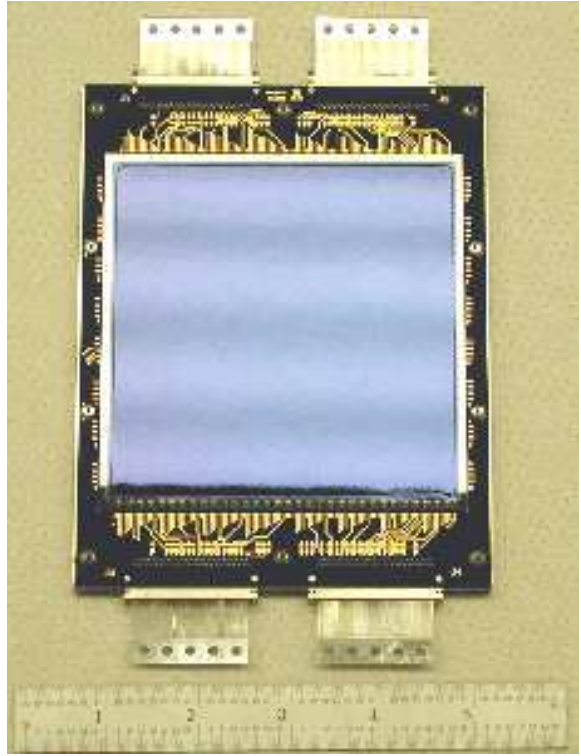}
   \end{tabular}
   \end{center}
   \caption[example] 
   { \label{fig:example} 
     The STA1600 chip after packaging.  The top and bottom areas
     connect to 8 outputs each. The photo-sensitive area is
     95 mm $\times$ 95 mm.}
   \end{figure}

\subsection{Thinning}

The backside processing of the 10k CCD is similar to processing
smaller devices\cite{mike1}.  The wafers are first mechanically lapped to 
250 ${\mu}m$.  Gold stub bumps are applied to each bond pad and then the
wafer is diced.  The die is hybridized to a 1.4 mm thick silicon
substrate with indium bumps matching the CCD bond pads.  Epoxy is used
as an underfill material.  Backside thinning is accomplished in a
selective etch which stops at the epitaxial layer.  A final etch
polishes the surface.  Backside coatings are applied using the
University of Arizona Chemisorption Charging process\cite{mike2}.  
A custom invar package and circuit board set has been designed and 
fabricated for the backside parts.  After packaging and wire bonding 
the device is ready for testing\cite{mike3}.
A mechanical flattness is achieved to support an f/4.5 beam of the
URAT instrument.

\section{Camera and Performance}

In addition to the thinned, backside-illuminated, science-grade STA1600
CCD detector, the 10k camera consist of a custom dewar, filter, shutter,
electronics and required interfaces.
The filter, made by Andover Corporation, is 12 mm thick, ultra-flat,
and has a diameter of 160 mm for a 683 to 747 nm bandpass.
The bandpass has been chosen to be as red as the astrograph
lens supports but to exclude the $H_{\alpha}$ region of the spectrum,
in order to avoid photons from emission nebulae on exposures taken for 
high accuracy centroiding of stellar images.
The filter is fixed mounted as dewar window and the separation of
the backside of the filter to the focal plane is only 5 mm.
In addition to the interference layers and coatings of that filter
a small (1.2 mm diameter) neutral density spot with a factor of
about 200 attenuation has been added near the center of the filter.
This will allow astrometric observations of bright stars in reference to
much fainter stars in the same field of view.

Figure 3 shows the dewar with an engineering grade STA1600 chip and
a clear glass window for testing.
This 10k camera is currently limited by the Leach electronics.  
A complete readout of the full 111 megapixel image requires 16 seconds. 
We are using an Astronomical Research GenIII camera with 2
ARC48 8-channel A-D boards.   The camera is running the CCD with a 912 kHz
serial clock and a 30 kHz parallel clock.  An ARC42 fiber optic timing board
relays output data from the ARC controller to the PC.  We are digitizing 16
bits for each of the 16 outputs.  This data rate of 912 kHz is limited
by the capacity of the current fiber optic card.  
At 912 kHz we achieve a read noise of 6 electrons RMS on a thinned 
STA1600 CCD.
Alternative electronics has been developed at STA and a readout at
20 MHz has been demonstrated on a frontside device.

A 150 mm aperture shutter was custom built for the 10k camera by
the Bonn instrumentation group\cite{bonn_shutter} (Fig.~4).
The camera is run by a Linux PC.
Operation of the shutter has been integrated and a completely
new interface to an upgraded astrograph is in preparation,
all controlled by the same PC from a command-line interface
suitable for robotic operation.
Image data files will be stored in a compressed FITS format,
about 120 MB per full-frame.

   \begin{figure}
   \begin{center}
   \begin{tabular}{c}
   \includegraphics[height=9cm]{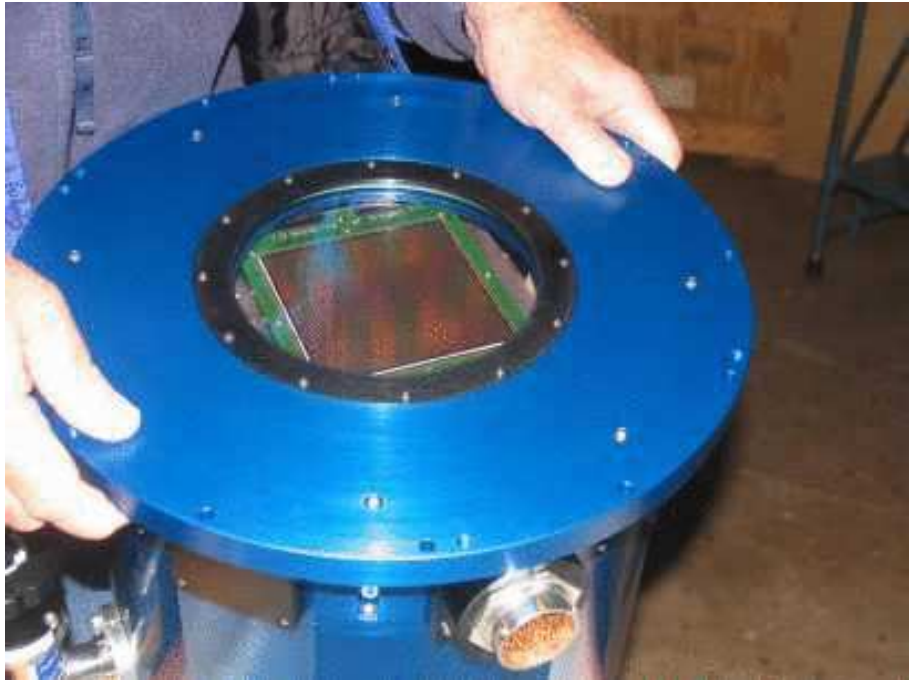}
   \end{tabular}
   \end{center}
   \caption[example] 
   { \label{fig:dewar} 
     The STA1600 chip inside the 10k camera dewar.
     Here an operational front-side chip is used.}
   \end{figure}

   \begin{figure}
   \begin{center}
   \begin{tabular}{c}
   \includegraphics[height=10cm]{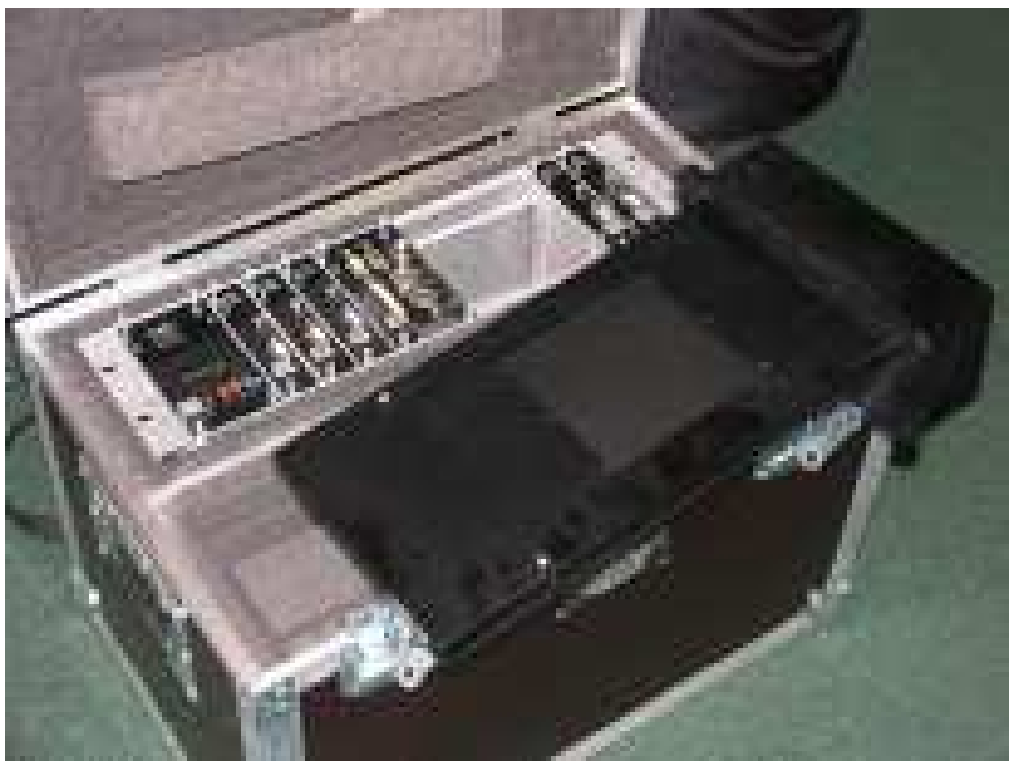}
   \end{tabular}
   \end{center}
   \caption[example] 
   { \label{fig:shutter} 
     The ``Bonn" shutter system for the 10k camera.
     This custom made shutter (black device) comes with
     control electronics to fit into a standard rack.}
   \end{figure}

\section{APPLICATIONS}

\subsection{Ground-based Star Catalogs}

The main application for this detector and camera is to support
DoD needs and requirements for star positions.  
This will also serve the general astronomical community by
providing highly accurate positions and proper motions of
millions of stars.

For star tracker applications (bright stars) the goal is to improve 
upon the Hipparcos Catalog\cite{hip97} positions, which have
steadily degraded due to accumulation of proper motion errors since 
their mean observing epoch in 1991.
This improvement can be accomplished by observing bright targets 
with the USNO astrograph and the new 10k camera through the neutral 
density spot on its filter.  Tycho-2 stars in the same 2.5 by 2.5 
degree field of view will serve as reference frame.

For Space Situational Awareness research the 10k camera 
can be used at either the astrograph or the URAT to determine
accurate positions of faint stars (down to R magnitude 18 and 21,
respectively).  For this application maximal sky coverage per exposure 
is needed.  
Figure 5 shows a focal plane layout with 4 of the 10k CCD detectors 
in their current packaging.  The circle is 333 mm in diameter, 
close to the limit of the astrograph focal plane area.
Attached to the astrograph this would provide 27 square degrees 
sky coverage in a single exposure.  This layout would need to be 
modified to be able to mount the 10k chips closer together for 
the slightly smaller URAT focal plane.

   \begin{figure}
   \begin{center}
   \begin{tabular}{c}
   \includegraphics[height=11cm,angle=-90]{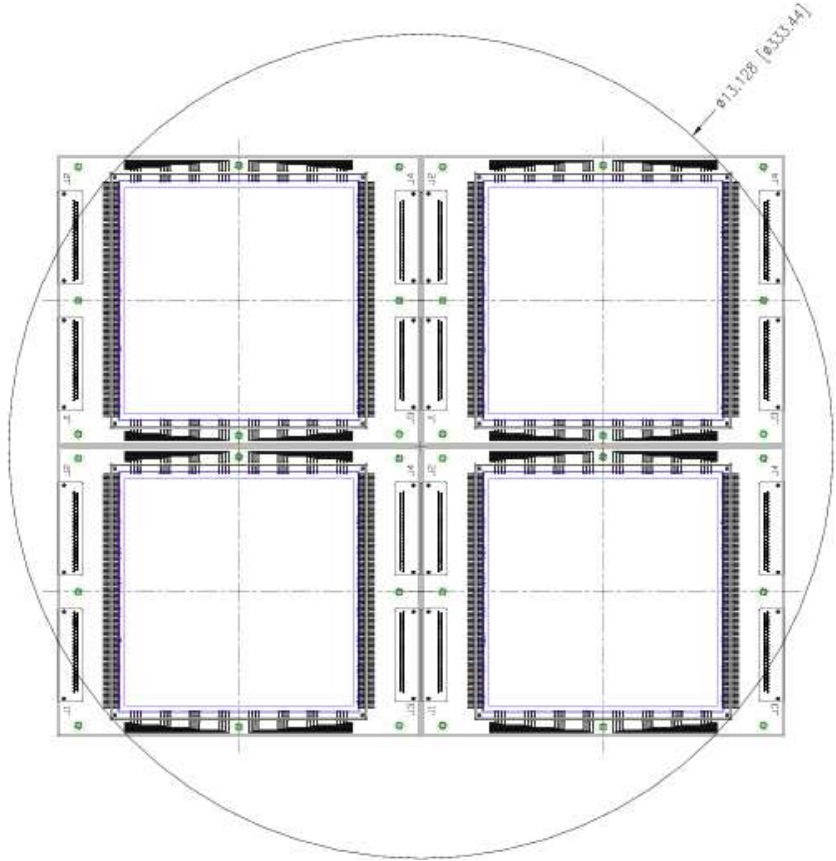}
   \end{tabular}
   \end{center}
   \caption[example] 
   { \label{fig:4shooter} 
     Layout of a 4-shooter focal plane based on the existing 
     STA1600 chip design.  This design would barely fit the
     astrograph field of view and is slightly too large for
     the URAT focal plane. A modification of the packaging
     is planned.}
   \end{figure} 

We plan to purchase such a ``4-shooter" camera in 2008.
This would allow us to construct a URAT focal plane and use it
at the astrograph.  After only 2 years of observing time from
the Cerro Tololo Inter-American Observatory (CTIO),
positions and parallaxes of stars in the 11 to 16 mag range
on the 5 to 10 mas level could be produced, significantly
improving current star catalog data.

If the USNO-lead Milli-Arcsecond Pathfinder Survey (MAPS)
mission\cite{jmaps,stare} is approved, the tie of the resulting
new celestial reference frame to fixed, extragalactic sources
would be performed by URAT, and funding is expected for the new, 
dedicated, ground-based telescope utilizing the 4-shooter camera 
based on an anti-blooming modified version of the STA1600 chip.

\subsection{In Space}

The initial design for the STA1600 called for 16 2-stage outputs that
could run at up to 15 MHz, resulting in a maximum frame rate of
approximately 2 frames per second (fps) with a resultant post-CDS read 
noise of 40 to 50 e$^{-}$ RMS.  Because it was designed to be operated 
at ground-based observatories, the current design has no built-in 
radiation mitigation capabilities.

A space-based implementation of this CCD could include an increase in 
the frame rate and improved radiation hardening.  In order to increase 
frame rate, the number of readout amplifiers would be increased to 32 
and 3-stage amps would be used rather than 2, increasing the speed to 
40 MHz per channel.
This approach would allow the frame rate to be increased from 2 to 
10 fps, with an increase in read noise to around 60 e$^{-}$ RMS. 
In order to improve radiation hardness, perhaps the most 
straightforward approach would be to
use p-channel rather than n-channel material.  
Numerous results\cite{marshall} have shown an increase of 
approximately an order of magnitude in hardness vs.~displacement 
damage is achieved when using p-channel material.  
Other solutions, such as active circuitry, could also be considered,
although these methods are less attractive due to their added 
complexity and potential negative impact on yield.

\section{DISCLAIMER}

Although some manufacturers are identified for the purpose of scientific
clarity, the USNO does not endorse any commercial product nor does the
USNO permit any use of this document for marketing or advertising.  We
further caution the reader that the equipment quality described here may
not be characteristic of similar equipment maintained at other
laboratories, nor of equipment currently marketed by any commercial
vendor.

\vspace*{5mm}

\acknowledgments   
 
We wish to thank the Office of Naval Research (ONR) for funding
this large, monolithic detector research program through a
Navy SBIR program and Sean Urban, Head of the Nautical Almanac
Office, for finding that sponsor.



\vspace*{5mm}

\begin{thebibliography}{1}

\bibitem{ucac2}
N.~Zacharias, S.~E.~Urban, M.~I.~Zacharias, G.~L.~Wycoff,
D.~M.~Hall, D.~G.~Monet, and T.~J.~Rafferty,
  ``The Second US Naval Observatory CCD Astrograph Catalog (UCAC2),"
  {\em AJ}, {\bf 127}, p.3043, 2004

\bibitem{redlens}
D.~Vukobratovich, T.~M.~Valente, R.~R.~Shannon, R.~A.~Hooker,
and R.~E.~Sumner, 
  {\em SPIE}, {\bf 1752}, p.245, 1991 

\bibitem{potsdam}
N.~Zacharias,
  ``Astrometric reference stars: from UCAC to URAT,"
  in {\em 3rd Potsdam Thinkshop on robotic telescopes},
  {\em AN}, {\bf 325}, p.631, 2004

\bibitem{uratd}
U.~Laux, and N.~Zacharias,
  ``URAT optical design options and astrometric performance,"
  in {\em Astrometry in the Age of the Next Generation of Large Telescopes},
  K.~P.~Seidelman and A.~K.~B.~Monet eds., {\em APS Conf.Ser.,} {\bf 338}
  p.106, 2005

\bibitem{uratp}
N.~Zacharias,
  ``The URAT Project,"
  in {\em Astrometry in the Age of the Next Generation of Large Telescopes},
  K.~P.~Seidelman and A.~K.~B.~Monet eds., {\em APS Conf.Ser.,} {\bf 338}
  p.98, 2005

\bibitem{111mpx}
R.~Bredthauer, K.~Boggs, G.~Bredthauer,
  ``A monolithic 111-M pixel high speed, high resolution CCD,"
  in {\em 2007 Image Sensor Workshop}, p.170, 2007

\bibitem{mike1}
M.~P.~Lesser,  
  ``Very Large Format Back Illuminated CCDs," 
  in {\em Scientific Detectors for Astronomy},  
  P.~Amico, J.~W.~Beletic, and J.~Beletic eds, p.137,  
  Kluwer Academic Publishers, 2004

\bibitem{mike2}
M.~P.~Lesser, V.~Iyer, 
   ``Enhancing Back Illuminated Performance of Astronomical CCDs,"
   {\em SPIE}, {\bf 3355}, p.23, 1998

\bibitem{mike3}
M.~P.~Lesser, D.~Ouellette, 
   ``Development of hybridized focal plane technologies,"
   {\em SPIE} {\bf 6276}, p.03, 2006

\bibitem{bonn_shutter}
K.~Reif,
  ``USNO150 a Bonn Shutter for the USNO Robotic Astrometric Telescope,"
   {\em user manual}, Bonn Instrumentation Group, 2006

\bibitem{hip97}
ESA, ``The Hipparcos Catalgoue," {\em European Space Agency}
  {\bf SP-1200}, 1997

\bibitem{jmaps}
B.~Dorland, N.~Zacharias, R.~Gaume, K.~J.~Johnston,
G.~Hennessy, V.~Makarov, and C.~Rollins,
  ``The Milli-Arcsecond Pathfinder Survey (MAPS) mission,"
  {\em BAAS}, {\bf 207}, abstract 117.04, 2005   

\bibitem{stare}
N.~Zacharias and B.~Dorland,
  ``Concept of a stare-mode astrometric mission,"
  {\em PASP}, {\bf 118}, p.1419

\bibitem{marshall}
C.~J.~Marshall, et al.,
  {\em SPIE}, {\bf 5499}, p.542, 2004

\end{thebibliography}
\end{document}